\documentclass{aa}

\def\amin{\ifmmode ^{\prime}\else$^{\prime}$\fi}
\newbox\grsign \setbox\grsign=\hbox{$>$}
\newdimen\grdimen \grdimen=\ht\grsign
\newbox\laxbox \newbox\gaxbox
\setbox\gaxbox=\hbox{\raise.5ex\hbox{$>$}\llap
     {\lower.5ex\hbox{$\sim$}}}\ht1=\grdimen\dp1=0pt
\setbox\laxbox=\hbox{\raise.5ex\hbox{$<$}\llap
     {\lower.5ex\hbox{$\sim$}}}\ht2=\grdimen\dp2=0pt

\def\lax{\mathrel{\copy\laxbox}}

\begin{document}

\title{On Be Star Candidates and Possible
Blue Pre-Main Sequence Objects
in the Small Magellanic Cloud}

\author{R.E.\ Mennickent \inst{1}
G.\ Pietrzy{\'n}ski \inst{1}\fnmsep\inst{2}
W.\ Gieren \inst{1}
O. Szewczyk\inst{2}
}

\institute{Universidad de Concepci{\'o}n, Departamento de F\'{\i}sica,
 Casilla 160--C, Concepci{\'o}n, Chile\\
\email{rmennick@stars.cfm.udec.cl, pietrzyn@hubble.cfm.udec.cl, wgieren@coma.cfm.udec.cl}
\and
Warsaw University Observatory, Al. Ujazdowskie 4,00-478, Warsaw, Poland\\
\email{szewczyk@astrouw.edu.pl}}

\date{}

\abstract{Recently the OGLE experiment has provided
accurate light curves and colours for about 2 millions stars in the Small
Magellanic Cloud.
We have examined this database for its content of Be stars, applying some selection
criteria, and we have found a sample of $\sim$ 1000
candidates. Some of these stars show beautiful light curves
with amazing variations
never observed
in any Galactic variable. We find outbursts in 13\% of the sample  (Type-1 stars),
high and low states in 15\%,
periodic variations in 7\%,
and the usual variations seen in Galactic Be stars
in 65\% of the cases.  The Galactic counterparts of Type-1
objects could be the outbursting Be stars found by
Hubert \& Floquet (1998) after the analysis of Hipparcos photometry.
We discuss the possibility that Type-1 stars could correspond
to Be stars with accreting
white dwarf companions
or alternatively,  blue pre-main sequence stars surrounded by thermally
unstable accretion disks.
We provide coordinates and basic photometric information for these stars
and some examples of light curves.
}

\authorrunning{R.E. Mennickent G. Pietrzynski W. Gieren O. Szewczyk}
\titlerunning{Be star candidates in the SMC}
\maketitle

\section{Introduction}
About 1/6 - 1/3 of all Galactic B-type stars with luminosity class
III-V once showed H$\alpha$ in emission and have therefore been
classified as Be
stars (e.g. Zorec \& Briot 1997). The optical
and near-infrared spectra of Galactic Be
stars usually display hydrogen, He I and singly
ionized metallic emission lines, whose double-peaked shape indicate their origin in a circumstellar
gaseous quasi-Keplerian disk.
Detailed  studies of Be stars in the Small Magellanic Cloud (SMC) have been
performed only in recent years, being especially
confined to open clusters like
NGC 330 (e.g. Keller, Wood and Bessell 1999).
These studies show the importance of studying Be stars in
the low metallicity environment of the SMC
since they serve as probes to test for the mechanisms of
disk formation and of global disk oscillations
(Baade et al.\ 2001, Hummel et al.\ 1999).

Over the past years, the microlensing projects (OGLE, MACHO, EROS)
have monitored millions of stars in the Magellanic Clouds
and Galactic bulge for variability. The huge resulting photometric databases are very well
suited not only for microlensing studies but also for many other issues of modern
astrophysics, including the distance scale, variable stars, star clusters etc.
In particular, the OGLE II project (Udalski, Kubiak and Szymanski, 1997),
has provided
accurate BVI measurements for about 6.5 million stars from the central
parts of the Magellanic Clouds (Udalski et al.\ 1998, 2000). Based on this same
material, a unique catalog containing about 68.000 variable stars has just been
released ({\.Z}ebru{\'n} et al.\ 2001).

The aim of this paper is to search the OGLE II  database
for Be star candidates and provide a much larger sample of these
stars in the SMC than previously known. This will provide the basis for
posterior follow-up spectroscopic studies of these objects, necessary to
fully understand their nature. We will also call attention
to several apparently new classes of light curves which appeared during our search
for Be stars in the SMC.

The paper has the following structure. Observational data and the
applied selection criteria are described in section 2.
Results are presented in section 3, and discussed in section 4.
In section 5, we present our current conclusions on the underlying nature of the observed types of variability.

\section{The Data}

During four years (1997-2000), about 4.5 and 2 million
stars were monitored in the LMC and SMC, respectively, in the course
of the OGLE II project (Udalski, Kubiak and Szymanski 1997).
Typically, about 400 I and 30 B and V observations were secured for
each star. An application of a newly developed photometric pipeline
(Wo{\'z}niak 2000, {\.Z}ebru{\'n} et al.\ 2001 ) to the data,
based on the DIA method introduced by Alard and Lupton (1998),
  resulted in the detection of about 68.000 variable stars
({\.Z}ebru{\'n} et al.\ 2002). This catalog, apart from presenting positions and light curves
for all its objects, provides in addition cross identifications
with objects detected during earlier reductions carried out with Dophot. It should
be stressed that the positions of the variable stars obtained with the DIA
package were obtained from the variable signal only, and may be in
principle quite different, especially in the very crowded regions,
from the positions obtained from the Dophot photometry. However, the great majority of
 variable sources detected with DIA were unambiguously  cross-correlated
with objects detected with Dophot ({\.Z}ebru{\'n} et al.\ 2002).

To carry out our search, we inspected the OGLE II SMC BVI maps,
and looked for
stars matching the usual luminosities and colours observed
in Galactic Be stars.
Absolute magnitudes of Be stars based on  Hipparcos parallaxes
(Wegner 2000) indicate that Be stars of luminosity classes IV-V and III
have absolute V-band magnitudes between
-4 and 0. However, Garmany \& Humphreys (1985) report on a population
of extremely luminous non-supergiant Be stars in the
Magellanic Clouds  with
absolute V-band magnitudes up to -6. Therefore,
we searched for stars having absolute V-band
magnitudes between -6 and 0. Assuming a distance modulus for the SMC
of 19.0 $\pm$ 0.2 mag, a visual
extinction of 0.3 mag and a typical amplitude of the photometric variations of Be stars
of 0.2 mag,
the above absolute magnitudes imply
apparent V-band magnitudes between 13.1 and 19.7. In addition to this magnitude
constraint,
we required that the selected stars
should have colours in the range in which Galactic Be stars are usually found,
viz.\, -0.4 $<$ B-V  $<$ 0.6 and -0.35 $<$
V-I $<$ 0.8.
Using these criteria, we obtained
a first sample of 5168 Be star candidates in our database.
As a next step, we extracted the I-band light curves for each of these objects
from the OGLE II catalog of variable stars ({\.Z}ebru{\'n} et al.\ 2002).
The light curves were visually inspected and
spurious variables, Cepheids and
eclipsing binaries rejected. We then proceeded in classifying the remaining 1056 stars according to
the appearance of their light curves. Our set of Be star candidates
is likely to be complete for stars showing a $rms$ larger
than 0.02 magnitudes in the I band.
However, we could have missed Be stars either not showing long-term
variations, or with  I-band amplitudes less than 0.02 magnitudes.

\section{Results}

Surprisingly, we found
not only a large sample of
685 objects showing light curves similar to those observed for Galactic Be stars
(we call them Type-4 stars, with stochastic or sometimes
quasi-periodic variations),
but also 139 objects showing outbursts (called Type-1 stars),
154 showing  high and low states (called Type-2 stars), and 78 stars
showing periodic or near-periodic long-term oscillations (Type-3 stars).
18 Type-1 stars also showed sudden luminosity jumps
like those observed in Type-2 stars. We called them Type-1/Type-2 stars.
In Figures 1-6, we show examples of light curves for objects of the Type-1
to Type-4 subclasses.
Basic information  (OGLE name, distance (in arcsec) to the nearest star detected with
Dophot, V magnitude, B-V and V-I colours, scatter of I-band magnitudes, and optional
comments) on these stars is given in  Tables 1-6.
It is striking that Type-2 and Type-3 light curves
have not been observed in any
Galactic variable, to the best of our knowledge, and  very few
cases of Galactic Type-1 stars seems to have been detected (see Section 3.1).
Therefore  all these stars clearly deserve further study.
In Fig.\,7 and Fig.\,8 we show the  V versus B-V, and V-I
versus B-V diagrams for the entire sample of stars.
It is observed that the number of Be star candidates declines
towards the extremes in both diagrams,
indicating that the ranges of magnitude and colour we
chose for our search were probably near-optimum to include all the Be stars in the database.
Regarding
the main sequence, most stars are displaced to the red
in the V versus B-V diagram; this is in principle
consistent with hydrogen
recombination (free-bound and free-free) emission
from circumstellar envelopes around Be stars
(Dachs, Engles and Kiehling 1988).
Keeping in mind that our data have not been
corrected for interstellar or circumstellar
reddening, the sharp blue cut-off in the B-V colour is notable.
This observational fact is consistent with the
vanishing of the Be phenomenon
in spectral types earlier than B0, partly due to the
efficiency of radiation pressure to remove circumstellar
gas in hotter stars.

\subsection{Type-1 stars:} The outbursts shown by Type-1 stars were basically
of two types:
those characterized by a sudden rise of luminosity
followed by a gradual decline, generally lasting tens of days with amplitudes
$\lax$ 0.2 mag, and
those with usually slower declines, lasting hundreds of days with amplitudes $>$ 0.2 mag, and
characterized by more symmetric rising and fading timescales.
In general, in both kinds of outbursts, the stars were redder when brighter
and we could not find any evidence of a correlation of
outburst rising delay  with  wavelength.
Some stars showed both kinds of outbursts, and sometimes it
was difficult to classify them when the
outburst was faint and not well resolved, or in the cases of noisy light curves.
For this reason we do not provide a sub-classification for Type-1
stars. Not always these stars are showing a constant quiescence brightness level.
For 16\% of Type-1 stars (22 objects) it was possible to identify
an outburst time-scale, i.e.\, a typical time for outburst recurrence, although this recursion  sometimes
was observed only as a transient phenomenon and not during the whole observing period. Basic
information for Type-1 stars is given in Tables 1 and 2. Exemplary light curves of this
type of variable stars are shown in Figs. 1 and 2.  The outburst
time-scales, given in Table 2, run from 16 to 332 days, with
a mean of 126 days. The time-scale histogram shows a decrease
of cases toward longer periods;
68\% of the objects show time-scales shorter than 150 days.
There is no clear correlation between outburst time-scale, and
stellar colour or apparent magnitude.

Our Type-1 objects probably correspond to the
``bumpers´´ found in the
Large Magellanic Cloud by Cook et al.\ (1995) after analysing the MACHO
collaboration database.
These authors report on the discovery of a group of bright
main sequence stars with V $\sim$ 15--17 which have a constant
luminosity for extended periods of time, but show occasional
outbursts at 10--30\% amplitude level. They found that the episodes in
more than 50\% of the cases are
asymmetric and shorter than about 50 days with a more rapid
brightening than dimming.
The fact that H$\alpha$ and H$\beta$ emission was observed in the
spectra of 7 of these stars led Cook et al.\ (1995)
to  suggest that these stars could be related to, or maybe identical to,
Galactic Be stars.

 The Galactic counterparts of Type-1 objects could be the outbursting
Be stars found by Hubert \& Floquet (1998) analyzing Hipparcos
photometry. They find short-lived outbursts in $\omega$ CMa and
other 13 Be stars, and also a long-lived outburst
in $\upsilon$ Cyg and another 7 Be stars. The outbursts were preferentially
detected in early Be stars with rather low
to moderate projected rotational velocities. The average
time scale for the recurrence of short-lived outburst is, however,
slightly larger than that observed for our SMC sample. Hubert \& Floquet (1998)
argue that both kinds of outbursts probably have the same origin and
can be explained by discrete high density emitting plasma events
seen under low inclination angles with respect to the rotational axis.
The number of cases presented by these authors is too small and
the photometry too scarce to make a detailed comparison with our
Type-1 stars, however.

Although at present we cannot
exclude the Be star nature of Type-1 objects,
we will present, in the next section,
two possible alternative  explanations for these objects.

\subsection{Type-2 stars and Type-1/Type-2 stars:}
Type-2 stars show sudden
brightness jumps of an amplitude of a few tenths of magnitude, with the star remaining at a
rather constant brightness level for hundreds of days. Type-2 stars usually don't show
additional variability, like the irregular variations observed in Type-4 stars,
except for rather smooth long-term changes.
Type-2 stars can be found in basically
the same colour range as Type-1 stars, but Type-2 stars are
statistically bluer than Type-1 stars.
 Eighteen Type-1 stars in our database also
showed high-low states like those observed in Type-2 stars.
These Type-1/Type-2 stars were only found in the range of 15.5~$<$~V~$<$~17 and in a narrow colour box characterized by
-0.2~$<$~B-V~$<$~0.2 and -0.2~$<$~V-I~$<$~0.25.  The existence of these ``transition" stars
supports the conjecture that
Type-1 and Type-2 stars could correspond to
the same kind of object. Basic information for
Type-2 and Type-1/Type-2 stars is given in
Tables 3 and 4.

\subsection{Type-3 stars:} In these periodic objects, we searched for periods longer than 0.2 days
using a variety of period searching algorithms implemented in
graphic C-routines  written by us, and also in IRAF. We found periodicities in the range of 17 to 720 days.
The accuracy for periods longer than 200 days is low, and these
variations can easily be quasi-periodic rather than strictly periodic.
51\% of the periods were found in the range of 17-100 days. The mean
period was 135 days. Many of the stars showed significant aliases in the
periodogram in addition to the main period, and 2 showed evidence of a variable period. Five stars
turned out to be eclipsing binaries, with evidence of additional variability not related
to changing aspects of the binary system.
In general, Type-3 stars were redder when brighter. All the relevant information is given in Table 5.
The 9 Type-3 stars in our sample with periods longer than 200 days  have
V-I colours in the range of 0.0-0.2 mag, and B-V colours in the range of
-0.1-0.0 mag.  They correspond
to stars with visual magnitudes between 15.5 and 18.0 mag.
In general, Type-3 stars show
similar colours than the other types of stars we are discussing, but they appear
more homogeneously distributed in Figs.\,7 and 8,
with a tendency to concentrate in the red
part of the colour-colour diagram.

\subsection{Type-4 stars:} They are the most numerous in our sample,
sharing colour ranges with the other types of variables.
They usually show stochastic
variability in time scales running from days to years  (Fig.\,6).
In some occasions we observed
quasi-periodic oscillations on time scales of weeks, and in others short-duration,
eclipse-like events. Information on Type-4 stars is given in Table 6.

\section{Discussion}

We have  reviewed the literature in order to find out if our stars
have already been detected as variable stars by other surveys.
In the original sample of 5168 objects, we found 8 stars classified
as Be stars by Keller et al.\ (1999). We classified
four of them as Type-4 (ngc330/219, ngc330/235, ngc346/259 and  ngc346/377),
and the others were missed by our visual inspection search method
since they show I-band $rms$ lower than 0.02 mag
(ngc330/238, ngc330/1054, ngc330/1064  and
ngc330/1239). We also found one Type-4 star
listed as extreme Be-star by
Garmany \& Humphreys (1985, their star AV433).  Another two
Type-4 stars are listed as Be/X-ray binaries and one as possible
Be/X-ray binary in the compilation by Haberl \& Sasaki (2000, their stars
20,23 and 6, respectively).
The fact that eight Type-4 stars have been previously classified as
Be stars or Be/X-ray binaries suggests that all Type-4 stars might
be related with Be stars in some way. Their light curve morphology,
quite similar to Galactic Be stars, supports this conjecture.


Type-1 light curves are reminiscent of those observed in FU Orionis stars,
thought to be caused by accretion events in protostellar disks around
young solar-type stars.
There is a notable similarity between the shape of the sharp
Type-1 outbursts with those observed in the extreme classical T Tauri
star EX Lupi (Herbig et al.\ 2001)
and the FU Ori star V 1057 Cyg (Bell et al.\ 1995).
However, the amplitude and duration of the outbursts are respectively
factors 20 and 1000 smaller in Type-1 stars.
On the other hand, the general appearance of Type-2 light curves, i.e.,
a rapid rise of luminosity followed by a "plateau"
of rather constant brightness resemble (except for a scale factor)
the initial stages of the outbursts observed in V1515 Cyg and
FU Ori (Bell et al.\ 1995).  Type-1 light curves are also quite similar
in shape to those observed in cataclysmic variables (CVs) of the dwarf novae
type. However, these objects experience in general  much larger
amplitude outbursts ($\sim$ 2-7 mags), have a rather constant
luminosity at maximum and usually return to the same quiescence
level after outburst.

The outbursts observed in  CVs and FU Ori stars have been modeled by
mass transfer instabilities in accretion disks (Meyer \&
Meyer-Hofmeister 1981, Bell et al.\ 1995).
A heating front propagates across the disk after a certain critical
density is reached, pushing  the disk into a hot state with
high mass transfer rate and consequently high accretion luminosity.
Rapid outburst rising times can be produced by outbursts starting
in the outer disk and propagating inwards,
and slower rising times can be produced
by outbursts triggered in the inner disk and propagating outwards.
We can speculate that the same phenomenon responsible for
FU Ori outbursts is also present in more massive young stars,
producing light curve patterns like the ones
observed in Type-1 and Type-2 stars.
In this view, the smaller outburst amplitude and shorter duration of Type-1 outburst
when compared with those observed in FU Ori stars
should correspond to accretion of matter
with lower mass transfer rates.
{\it Sharp and hump-like outbursts should correspond to outside-in and inside-out
outbursts, respectively}. Type-2 stars could be explained by
outside-in outbursts followed by a sustained period of a high mass-transfer rate.
A characteristic of the FU Ori outbursts is their larger amplitude at
shorter wavelengths,
contrary to what is
observed in Type-1 stars. However, dwarf novae are redder at maximum.

In connection with the above ideas, it is interesting that one of our
objects classified as Type-1 was identified by Beaulieu et al.\
(2001) as a possible pre-main sequence star (ESHC2).
In principle, this is
in contradiction with the  Cook et al.\ (1995) classification of some
of these objects as Be stars.
It is possible that the presence of Balmer emission not
necessarily implies a Be star classification for the
``bumpers" detected in the LMC by  Cook and collaborators.
In fact, the low ionization metal absorption lines usually
found in optical spectra of Herbig Ae/Be stars could be hard
to detect in low metallicity stellar photospheres like those found
in the SMC and LMC.  Therefore, it is reasonable to ask if perhaps
{\it all} of Type-1 and  Type-2 objects are actually pre-main sequence stars
with FU Ori-like accretion disk activity.   In this case, the mass
accretion rates need to be much higher than those observed in normal
Be star discs, or even in Be/X-ray binaries, to yield
luminosities comparable or larger than the luminosity of
the accreting B-type star.
Since the nebulosities defining Herbig Ae/Be objects
cannot be resolved at the distance of the Magellanic Clouds,
the answer to  the above
question needs to await future
spectroscopic studies, especially in the infrared, and possibly
very high-resolution
imaging studies which could reveal the surrounding nebulosities.
We must keep in mind,
however, that Hipparcos photometry of
Galactic Herbig Ae/Be stars apparently does not reveal Type-1
like activity (van den Ancker et al.\,1998). Also, the optical
light of some Galactic Be stars showing outbursts,
like $\lambda$ Eri and $\mu$ Cen, is apparently
dominated by the stellar continuum during quiescence,
and the spectrum does not reveal the strong emission lines typical of
massive Herbig Ae/Be envelopes.
Although it is not clear if these Galactic Be stars are the same
kind of object that Type-1/Type-2 stars,
we recognize that these two facts are the
weak points in the pre-main sequence scenario. We therefore offer
an alternative explanation in the next paragraph.

Type-1 star outbursts
resemble also in shape the optical outbursts observed in some
Be/X-ray binaries (e.g. A 0538-66, Densham et al.\, 1983),
except for their much smaller amplitudes and redder colours. It is
currently assumed that
in these objects a neutron star companion
undergoes episodes of mass accretion when passing through the envelope
of a Be star primary (e.g., Okazaki \& Negueruela 2001).
These high-energy phenomena are no doubt relevant in the X-ray domain,
but should not show up at optical and near infrared wavelengths.
In addition, the optical counterparts of Galactic and Magellanic Cloud
Be/X-ray binaries have spectral types earlier than B2 (Negueruela 1998).
This is in principle
contrary to the distribution of B-V colours in Fig.\,7.
However, the similarity between the {\it X-ray} Type-I and Type-II
outbursts observed in Be/X-ray binaries and the narrow and hump-like
outburst observed in Type-1 stars
suggest that they could be due to accretion by a {\it white dwarf},
not a neutron star. The Be+WD system
should produce less energetic outburst, eventually much prominent in
the optical and infrared. Favouring this hypothesis  is the fact
that many Be+WD binaries, even with late-type B stars,
are expected from evolutionary calculations (Waters et al.\,1989).
Only few candidates for these systems have been found up to now,
likely due that the white dwarfs, if present, should be hard to
detect (Apparao 1991).
The hypothesis of accreting white dwarf companions  would
naturally explain some of the Type-1 light curves. The same model
for periodic X-ray outbursts in Be/X-ray binaries by
Okazaki \& Negueruela (2001) could eventually be applied to Be+WD systems.
In this model, {\it periodic narrow outbursts occur in systems with high
orbital eccentricity, every time that the compact object captures gas
from the disc at every periastron passage. Hump-like outbursts occurs
in systems with low eccentricity, when the outer disc eventually
passes beyond the orbit of the companion}. Accordingly to Apparao (1991),
observations of time delays between enhancements of optical line and
continuum emission could identify Be stars with white dwarf companions.
We must keep in mind,  that
only a small fraction of Type-1 objects shows periodic or quasiperiodic
outbursts, however.

Concerning Type-3 stars, they can be ruled out to be Cepheids, since
their large periods (sometimes hundreds of days) are incompatible
with their inferred luminosities. On the other hand,
the periods are  one or two orders of magnitude longer than those
observed in Galactic pulsating B stars. This hints at
the possibility that Type-3
variability is linked to the envelope rather than to the
star itself.  There is evidence for global
one-armed oscillations in the envelopes
of Galactic Be stars (Okazaki 1997),
but the associated photometric variations
are not so strictly periodic as those observed in Type-3 stars,
the time-scales are longer, about 7 years in average,
and, from a theoretical point of view, the oscillations are
expected to disappear in the low metallicity
environment of the SMC or, alternatively, to slow down, yielding
periods in excess of several years (Hummel et al.\ 2001), in
disagreement with the observed Type-3 oscillations. The
distribution of Type-3 stars in the colour-colour diagram of Fig.\,8
suggests that these stars are perhaps not be linked to the Be star
phenomenon at all.  Preliminary spectroscopy of
one Type-3 object seems to corroborate this view. It
shows double  H$\alpha$ emission with a deep central absorption and
total equivalent width $\sim$ 100\AA, too large for an isolated Be star
(Mennickent et al.\, in preparation). We speculate that Type-3
stars could be young objects with massive envelopes ongoing
some kind of oscillation.

\section{Conclusions}

We have presented a sample of $\sim$ 1000 Be star
candidates in the Small Magellanic Cloud, giving positions
and basic photometric information along with periodicities when present.
Many of these objects show apparently new kinds of photometric
variability never seen before, and this feature enabled us to carry out
an empirical classification based on the light curve appearance.
There is evidence that some of these objects are truly
Be stars, but spectroscopy is needed to
confirm this suspicion. On the other hand, the hypothesis
that  at least part of the Type-1 and Type-2 stars are
Be-stars with accreting white dwarfs, or alternatively,
pre-main sequence stars
showing accretion disk instabilities,
should be
studied more thoroughly.

The understanding of the phenomena causing
the photometric variability of the stars
discussed in this paper
requires knowledge of the spectral energy distributions of these stars.
We plan  to discuss, in a forthcoming paper, the detailed
photometric properties of these objects
along with spectroscopic data.  It should be interesting
to test the accretion scenario for Type-1 stars. Since
the theory of accretion is rather well developed, it could be
a source of physical information about the environments
surrounding these objects. We also plan
to search the OGLE-II database for
Be stars in the Large Magellanic Cloud.
It should be interesting
to investigate if the same types of light curves are observed
in the LMC, whose metallicity is
intermediate between the SMC and our Galaxy,
and how the light curve properties
are modified by the change in metallicity.

\begin{acknowledgements}

This work was supported by Grant Fondecyt 1000324 and
DI UdeC 202.011.030-1.0.
 
The authors are very grateful to Professor Bohdan Paczynski
for many interesting suggestions on this paper. We would also
like to thank the OGLE collaboration for making their data
public domain, thus enabling us to carry out this research.
We also acknowledge the anonymous referee who made valuable
comments to improve a first version of this manuscript.
\end{acknowledgements}

{\bf Figure Captions}

Fig.1 Examples of Type-1  Be star candidates showing sharp outbursts.

Fig.2 Examples of Type-1  Be star candidates showing hump-like outbursts.

Fig.3 Examples of Type-1/Type-2  Be star candidates 
  showing outbursts {\it and}
  high-low states.

Fig.4 Examples of Type-2  Be star candidates 
  showing high and low states.

Fig.5 Examples of Type-3  Be star candidates  showing periodic
  oscillations.

Fig.6 Examples of Type-4  Be star candidates  showing random
  variability.

Fig.7 The V vs.\ B-V diagram for the sample stars. Blue dots correspond to
  Type-1 stars, red dots to Type-2 stars, green dots to Type-3 stars,
  circles to Type-4 stars and black dots to Type-1 stars showing high and low
  states, as Type-2 stars. The track of the
  main sequence (Allen 2000) is shown for reference.
  Note the sharp cut-off at the blue edge.
  
Fig.8 The V-I vs.\ B-V diagram for the sample stars.
  The track of the main sequence  (Allen 2000) is shown as a reference.
  Meaning of the symbols are
  as in Fig.7.

\vspace*{10cm}
 
\begin{table}
\caption[]{Type-1 stars. $\Delta \Phi$ is the distance (in arcsec) to the
nearest star detected with Dophot. }
\begin{center}
\begin{tabular}{lccccc} \hline  \multicolumn{1}{c}{Star}&
\multicolumn{1}{c}{$\Delta\Phi$} & \multicolumn{1}{c}{V} &
\multicolumn{1}{c}{B-V}& \multicolumn{1}{c}{V-I} &
\multicolumn{1}{c}{$rms$} \\
\hline
003623.36-733922.3& 0.601& 16.158& -0.233& -0.212& 0.012 \\
003832.64-732234.9& 0.076& 15.264& -0.139& -0.117& 0.01 \\
003918.20-733656.6& 0.245& 14.548& 0.484& 0.714& 0.017 \\
003922.09-732531.6& 0.031& 16.931& -0.041& -0.146& 0.019 \\
004036.15-732921.6& 0.04& 16.548& 0.13& -0.03& 0.062 \\
004207.86-734501.9& 0.136& 16.804& -0.055& 0.049& 0.02\\
004215.08-731710.4& 0.097& 15.966& -0.157& -0.055& 0.019\\
004231.84-732200.9& 0.024& 15.376& -0.106& -0.019& 0.077\\
004326.72-725910.3& 0.439& 16.858& -0.128& -0.127& 0.018\\
004502.43-732318.6& 0.137& 16.777& -0.048& 0.039& 0.012\\
004624.69-724657.3& 0.609& 19.333& 0.203& 0.463& 0.045\\
004624.74-731941.7& 0.102& 15.851& 0.069& 0.169& 0.09\\
004631.41-730335.6& 0.085& 17.113& -0.118& -0.002& 0.04\\
004646.81-731849.2& 0.054& 16.701& -0.013& 0.207& 0.015\\
004650.39-731017.7& 0.092& 15.513& -0.105& -0.022& 0.014\\
004653.17-732330.2& 0.327& 15.252& -0.033& 0.012& 0.014\\
004717.43-725227.6& 0.119& 17.071& 0.353& 0.403& 0.136\\
004757.42-731050.8& 0.122& 15.714& -0.036& 0.041& 0.02\\
004800.14-730728.6& 0.055& 16.354& -0.226& 0.046& 0.101\\
004803.08-725404.2& 0.347& 16.704& -0.127& -0.126& 0.015\\
\hline
\end{tabular}
\thanks{ The complete version of this table, as in tables 3, 5 and 6,
is in the electronic version of the Journal. }
\end{center}
\end{table}

\begin{table*}
\caption[]{Type-1 stars showing transient quasi-periodic outbursts. ART
means ``after removing trend", BHS
``before high state" and BF
``before fading".  }
\begin{center}
\begin{tabular}{lcccccc} \hline  \multicolumn{1}{c}{Star}&
\multicolumn{1}{c}{$\Delta\Phi$} & \multicolumn{1}{c}{V} &
\multicolumn{1}{c}{B-V}& \multicolumn{1}{c}{V-I} &
\multicolumn{1}{c}{$rms$}  & \multicolumn{1}{c}{Time-scale (days)}
\\
\hline
004957.09-730204.5& 1.776& 17.129& -0.133& 0.122& 0.049& 332,ART      \\
005039.05-725751.4& 0.028& 17.211& 0.15& 0.301& 0.078& 136          \\
005118.72-732846.3& 0.158& 15.765& 0.039& 0.248& 0.013& 216,ART\\
005149.34-724134.3& 0.069& 16.026& -0.013& 0.163& 0.02& 38,BF   \\
005157.18-730811.8& 0.183& 16.544& -0.058& 0.092& 0.025& 22,ART    \\
005235.60-723751.7& 0.03& 14.573& -0.145& -0.114& 0.045& 94      \\
005312.58-725533.6& 0.039& 16.024& -0.121& -0.029& 0.037& 58       \\
005321.08-724548.3& 0.095& 15.526& -0.201& -0.067& 0.049& 231        \\
005355.66-724359.2& 0.151& 15.56& -0.092& -0.057& 0.114& 85 or 42,ART,BHS \\
005623.53-723926.3& 0.071& 16.225& 0.065& 0.309& 0.022& 215,ART   \\
005758.52-722228.7& 0.084& 15.794& -0.088& 0.002& 0.03& 40,ART,BF      \\
005802.33-724137.5& 0.031& 17.092& -0.167& -0.154& 0.057& 16,ART      \\
005950.21-722817.4& 0.123& 17.342& -0.112& 0.03& 0.017& 173,ART         \\
010001.47-724046.9& 0.062& 15.876& -0.187& -0.039& 0.154& 133 or 101,ART \\
010023.52-723302.7& 0.025& 15.6& -0.194& -0.079& 0.022& 168,ART     \\
010120.64-721118.7& 0.097& 15.489& -0.065& 0.078& 0.019& 75,ART       \\
010332.59-720326.5& 0.086& 14.91& -0.207& -0.129& 0.146& 127,ART        \\
010409.79-723835.6& 0.086& 15.325& -0.097& 0.046& 0.026& 112,BHS          \\
010838.56-723633.5& 0.036& 16.115& -0.075& 0.092& 0.035& 136            \\
\hline \end{tabular}
\end{center} \end{table*}

\begin{table}
\centering
\vspace*{23 cm}
\end{table}

\begin{table*}
\caption[]{Type-2 stars.}
\begin{center}
\begin{tabular}{lccccc} \hline  \multicolumn{1}{c}{Star}&
\multicolumn{1}{c}{$\Delta\Phi$} & \multicolumn{1}{c}{V} &
\multicolumn{1}{c}{B-V}& \multicolumn{1}{c}{V-I} &
\multicolumn{1}{c}{$rms$}
\\
\hline
004234.39-733058.2& 0.043& 15.989& -0.148& 0.002& 0.021\\
004258.05-730230.7& 0.036& 16.059& 0.028& 0.176& 0.019\\
004319.46-731705.7& 0.099& 16.403& -0.056& 0.132& 0.032\\
004330.69-732034.2& 0.028& 16.428& -0.194& -0.177& 0.009\\
004414.25-731215.5& 0.086& 17.297& -0.107& -0.028& 0.023\\
004443.05-731316.5& 0.068& 16.217& -0.072& 0.028& 0.017\\
004504.35-724449.9& 0.076& 17.906& 0.336& 0.549& 0.051\\
004510.40-731648.1& 0.104& 15.855& -0.105& -0.022& 0.009\\
004539.31-731040.1& 0.433& 17.849& -0.104& 0.702& 0.088\\
004546.74-733233.1& 0.008& 16.284& -0.032& 0.158& 0.032\\
004556.84-730919.8& 0.068& 17.128& -0.057& 0.031& 0.013\\
004601.22-731821.9& 0.11& 16.374& 0.033& 0.255& 0.018\\
004610.97-732535.2& 0.037& 15.439& 0.009& 0.032& 0.146\\
004633.48-730352.1& 0.026& 15.489& -0.066& 0.063& 0.04\\
004639.77-725241.0& 0.047& 15.614& -0.179& -0.173& 0.011\\
004651.07-730215.3& 0.048& 15.801& -0.026& 0.176& 0.026\\
004655.32-731207.6& 0.094& 16.164& -0.113& 0.103& 0.03\\
004714.29-731044.0& 0.09& 17.635& -0.022& 0.043& 0.022\\
004714.53-731349.7& 0.131& 15.068& -0.196& -0.172& 0.009\\
004719.31-732533.6& 0.11& 16.449& -0.074& 0.145& 0.073\\
\hline \end{tabular}
\end{center} \end{table*}

\begin{table*}
\caption[]{Type-1/Type-2 stars. The nomenclature is as in Table 2.}
\begin{center}
\begin{tabular}{lcccccc} \hline  \multicolumn{1}{c}{Star}&
\multicolumn{1}{c}{$\Delta\Phi$} & \multicolumn{1}{c}{V} &
\multicolumn{1}{c}{B-V}& \multicolumn{1}{c}{V-I} &
\multicolumn{1}{c}{$rms$} &
\multicolumn{1}{c}{Time-scale (days)}
\\
\hline
004402.00-733129.4& 0.221& 15.456& -0.003& 0.232& 0.016 &284\\
004406.65-732938.2& 0.034& 16.285& -0.029& 0.114& 0.02 &\\
004650.21-732807.3& 0.042& 15.442& -0.192& -0.191& 0.017 &\\
004738.11-731126.1& 0.12& 15.484& -0.114& 0.022& 0.008&\\
004800.73-732253.0& 0.137& 15.842& -0.048& 0.144& 0.018&\\
004817.58-725028.8& 0.133& 16.329& -0.109& -0.048& 0.016&\\
004848.27-732611.3& 0.062& 15.462& -0.098& 0.016& 0.067&39,BF\\
004858.25-724119.4& 0.054& 16.157& -0.074& 0.076& 0.015&\\
005045.50-730112.9& 0.071& 16.398& -0.057& 0.038& 0.019&\\
005053.20-731030.8& 0.046& 15.925& -0.054& 0.157& 0.096&\\
005105.65-731311.5& 0.044& 15.911& -0.032& 0.208& 0.047&\\
005112.13-725656.2& 0.08& 16.196& -0.082& 0.09& 0.034&52,ART\\
005141.43-731129.9& 0.051& 16& -0.123& 0.024& 0.091&\\
005147.58-730924.7& 0.066& 15.629& -0.073& 0.077& 0.045&\\
005456.92-731200.4& 0.056& 16.344& -0.156& -0.143& 0.01&\\
005651.99-724027.1& 0.059& 15.735& -0.065& -0.015& 0.067&\\
005854.06-722841.9& 0.114& 17.014& -0.04& 0.134& 0.042&\\
010701.71-724754.9& 0.036& 15.601& -0.152& -0.053& 0.021&\\
 \hline \end{tabular}
\end{center} \end{table*}

\begin{table}
\centering
\vspace*{23 cm}
\end{table}

\begin{table*}
\caption[]{Type-3 stars. ART means ``after-removing-trend",
ChP changing period, LA low amplitude, MA
multiple aliases, and VA variable amplitude.}
\begin{center}
\begin{tabular}{lcccccc} \hline  \multicolumn{1}{c}{Star}&
\multicolumn{1}{c}{$\Delta\Phi$} & \multicolumn{1}{c}{V} &
\multicolumn{1}{c}{B-V}& \multicolumn{1}{c}{V-I} & \multicolumn{1}{c}{$rms$}
& \multicolumn{1}{c}{Period (days)} \\
\hline
003813.22-734144.7& 0.291& 16.607& -0.045& 0.022& 0.058& 187$\pm$19   \\
003832.82-734404.2& 0.385& 17.447& 0.448& 0.609& 0.106& 79$\pm$3,MA    \\
003833.35-731510.1& 0.177& 16.467& 0.019& 0.084& 0.064& 164$\pm$3         \\
003927.34-733309.6& 0.322& 15.855& -0.158& 0.057& 0.055& $\sim$27,ART,MA   \\
003952.15-730057.7& 0.53& 18.815& -0.042& 0.022& 0.077& 107$\pm$6,double eclipse?\\
004019.36-730402.9& 2.172& 16.207& -0.033& 0.146& 0.059& 200$\pm$22,VA             \\
004024.94-734416.1& 1.58& 16.531& -0.018& 0.04& 0.01& $\sim$448                       \\
004211.38-732437.7& 0.345& 17.916& -0.089& 0.01& 0.058& 74$\pm$3                        \\
004212.21-734120.0& 0.017& 15.921& -0.197& -0.138& 0.039& 46$\pm$1,alias 1.01932          \\
004323.84-733930.1& 0.636& 16.341& -0.19& -0.145& 0.016& 62$\pm$2,alias 0.98134             \\
004336.91-732637.7& 0.097& 14.178& 0.181& 0.331& 0.031& 118$\pm$7,MA                          \\
004357.18-732019.9& 0.139& 16.51& 0.386& 0.668& 0.065& 29$\pm$1    \\
004454.66-732802.9& 0.183& 14.817& 0.286& 0.576& 0.03& 17.37$\pm$0.15,VA\\
004517.55-732343.6& 0.044& 17.513& 0.306& 0.222& 0.094& 171$\pm$15        \\
004541.10-731219.2& 0.082& 17.149& 0.089& 0.24& 0.032& 182$\pm$18           \\
004541.80-724932.3& 0.046& 17.252& 0.308& 0.48& 0.206& 85$\pm$3               \\
004554.14-731404.3& 0.781& 15.483& 0.517& 0.709& 0.01& 99$\pm$4,LA              \\
004633.06-731919.1& 0.129& 17.15& -0.012& 0.181& 0.021& 38$\pm$1,ART              \\
004633.76-731204.3& 0.092& 14.058& 0.206& 0.385& 0.057& 184$\pm$18,eclipsing        \\
004652.03-731423.7& 0.084& 15.995& -0.069& 0.087& 0.021& $\sim$567,other 67?          \\
004653.24-724300.0& 0.495& 19.469& 0.03& 0.12& 0.166& 66$\pm$3                          \\
004702.75-730618.3& 0.126& 16.57& 0.122& 0.243& 0.016& 17.3$\pm$0.15,ART,eclipsing        \\
004723.53-730347.0& 0.11& 16.328& -0.043& 0.162& 0.016& 101$\pm$2,ART,eclipsing              \\
004748.20-731906.1& 0.073& 18.355& 0.194& 0.392& 0.079& 177$\pm$10                            \\
004750.14-731316.4& 0.222& 15.475& 0.269& 0.471& 0.009& 30.0$\pm$0.3,LA                          \\
004816.01-730635.7& 0.212& 15.917& 0.123& 0.338& 0.04& 26.45$\pm$0.20,strange phase curve,also Type-1?\\
004833.67-732955.6& 0.221& 19.403& 0.129& 0.417& 0.14& 161$\pm$12                                      \\
004843.23-731415.8& 0.111& 16.736& -0.081& 0.066& 0.062& $\sim$357,VA                                    \\
004849.06-724309.2& 0.088& 17.588& 0.446& 0.639& 0.108& 150$\pm$15,rather time-scale                       \\
\hline
\end{tabular}
\end{center}
\end{table*}

\begin{table*}
\caption[]{Type-4 stars.}
\begin{center}
\begin{tabular}{lccccc} \hline  \multicolumn{1}{c}{Star}&
\multicolumn{1}{c}{$\Delta\Phi$} & \multicolumn{1}{c}{V} &
\multicolumn{1}{c}{B-V}& \multicolumn{1}{c}{V-I} &
\multicolumn{1}{c}{$rms$} \\
\hline
003618.14-734624.6& 0.086& 14.929& 0.353& 0.779& 0.279\\
003618.15-734247.4& 1.419& 16.177& -0.131& -1E-3& 0.036\\
003619.44-732553.1& 0.132& 19.519& 0.258& 0.516& 0.182\\
003621.53-732610.6& 0.05& 14.741& 0.067& 0.214& 0.124\\
003623.02-734651.0& 0.016& 19.551& 0.484& 0.531& 0.304\\
003645.28-733259.7& 1.863& 15.59& -0.024& 0.2& 0.015\\
003652.74-732857.9& 3.222& 14.397& 0.081& 0.177& 0.016\\
003715.60-734546.0& 1.804& 16.09& 0.453& 0.647& 0.013\\
003726.27-731420.0& 0.511& 19.63& 0.318& 0.559& 0.25\\
003730.99-731112.4& 1.965& 15.846& 0.109& 0.209& 0.01\\
003744.41-730447.6& 0.109& 16.996& -0.087& -0.051& 0.028\\
003804.71-735150.5& 0.329& 19.434& 0.009& 0.092& 0.065\\
003810.16-730138.4& 1.22& 17.296& 0.59& 0.767& 0.02\\
003813.77-731340.0& 0.208& 17.335& -0.081& 0.116& 0.016\\
003817.01-734852.6& 0.025& 15.199& 0.462& 0.709& 0.28\\
003827.69-733314.5& 1.499& 16.126& -0.156& 0.124& 0.021\\
003833.29-732629.4& 1.317& 16.725& 0.414& 0.619& 0.011\\
003836.25-734813.4& 0.303& 15.383& -0.14& -0.082& 0.022\\
003837.66-731158.3& 0.093& 14.894& -0.085& 0.057& 0.028\\
\hline
\end{tabular}
\end{center}
\end{table*}

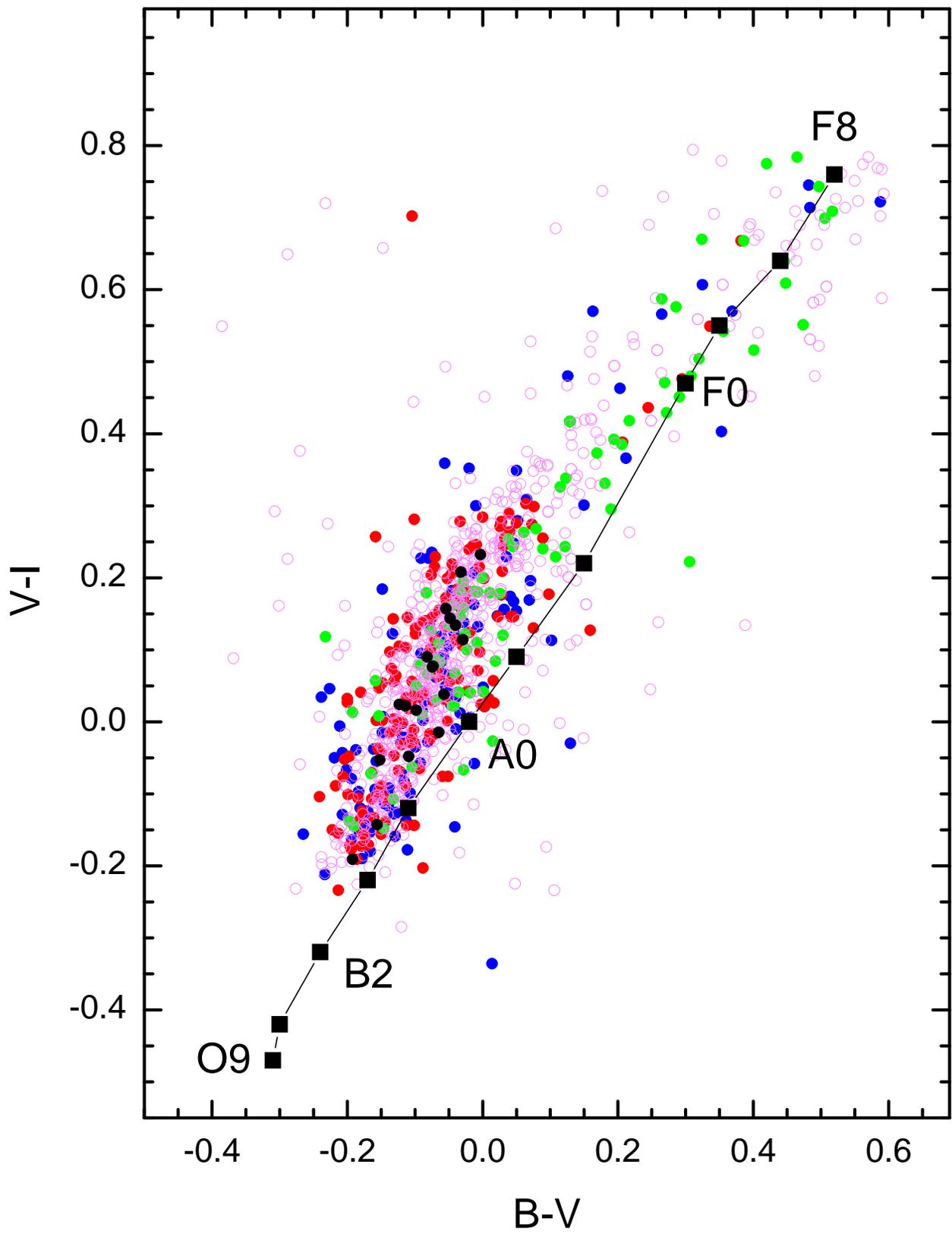
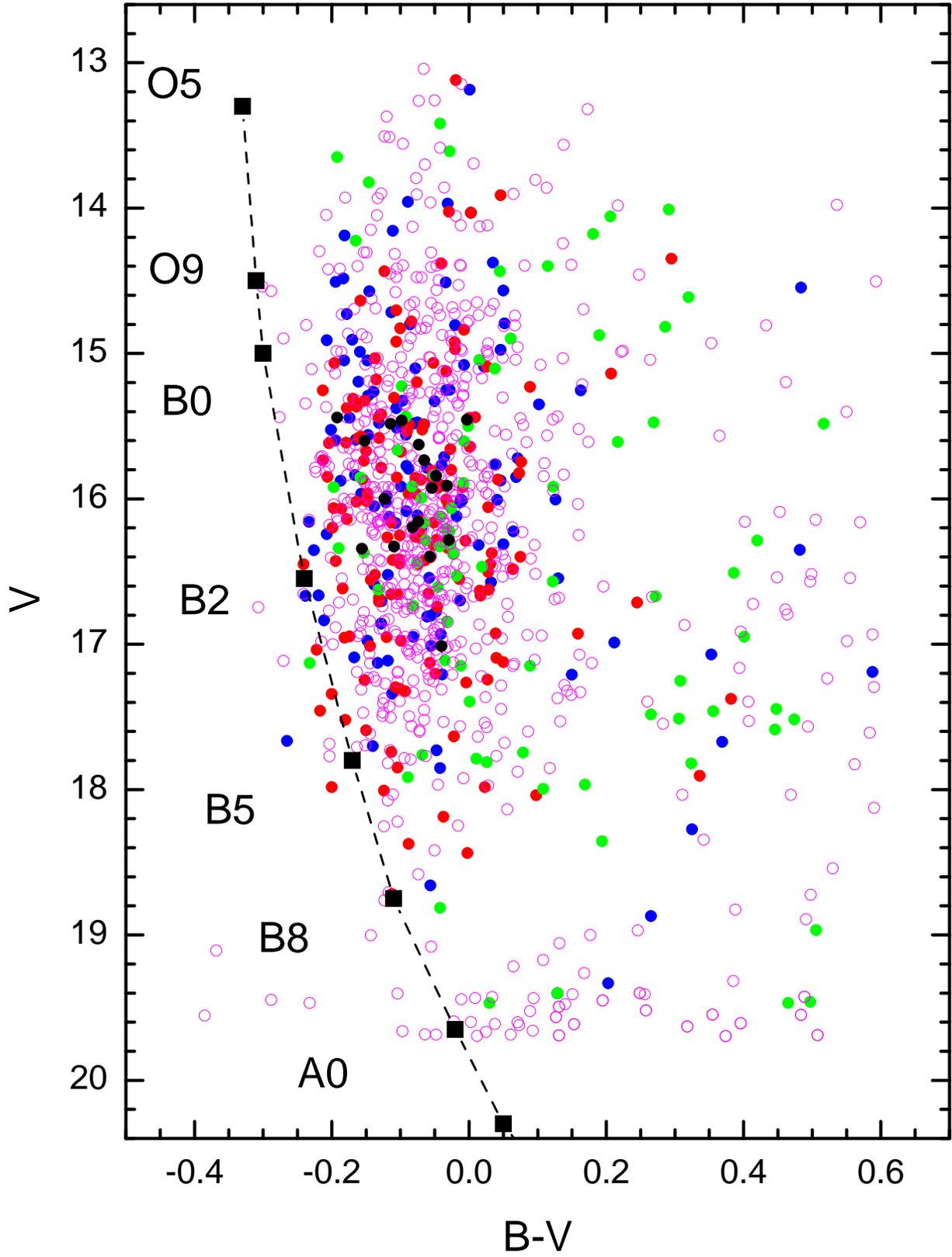
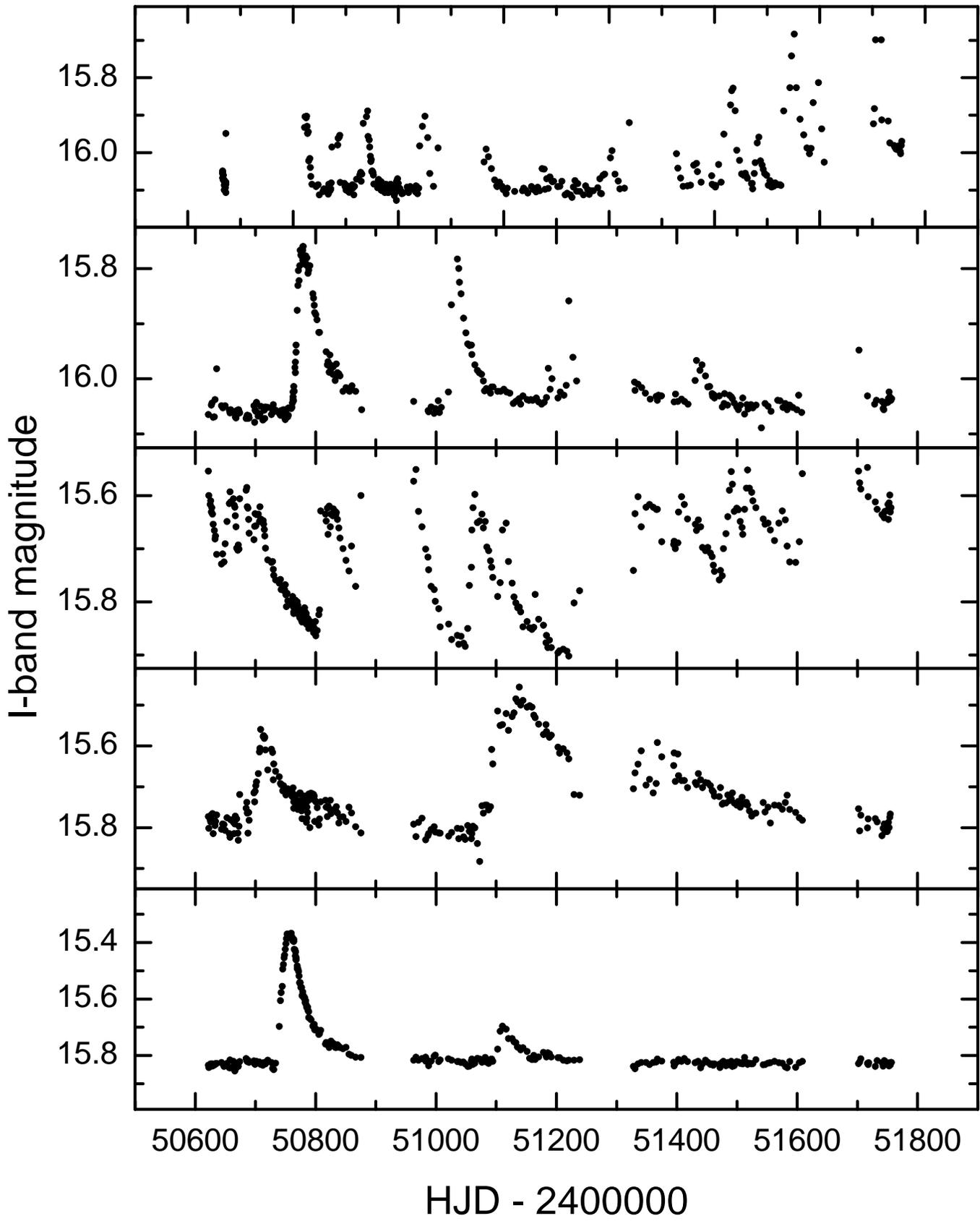
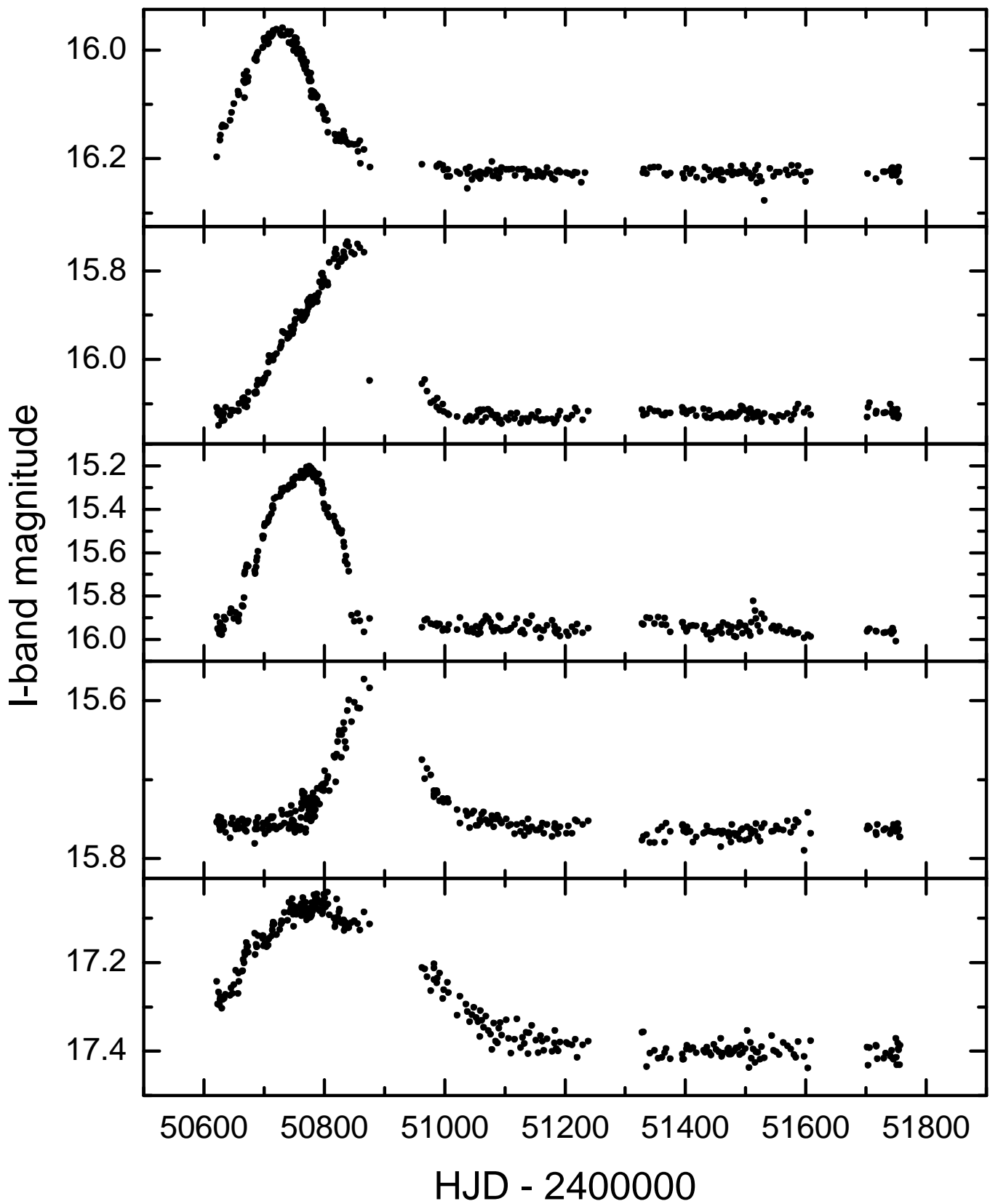
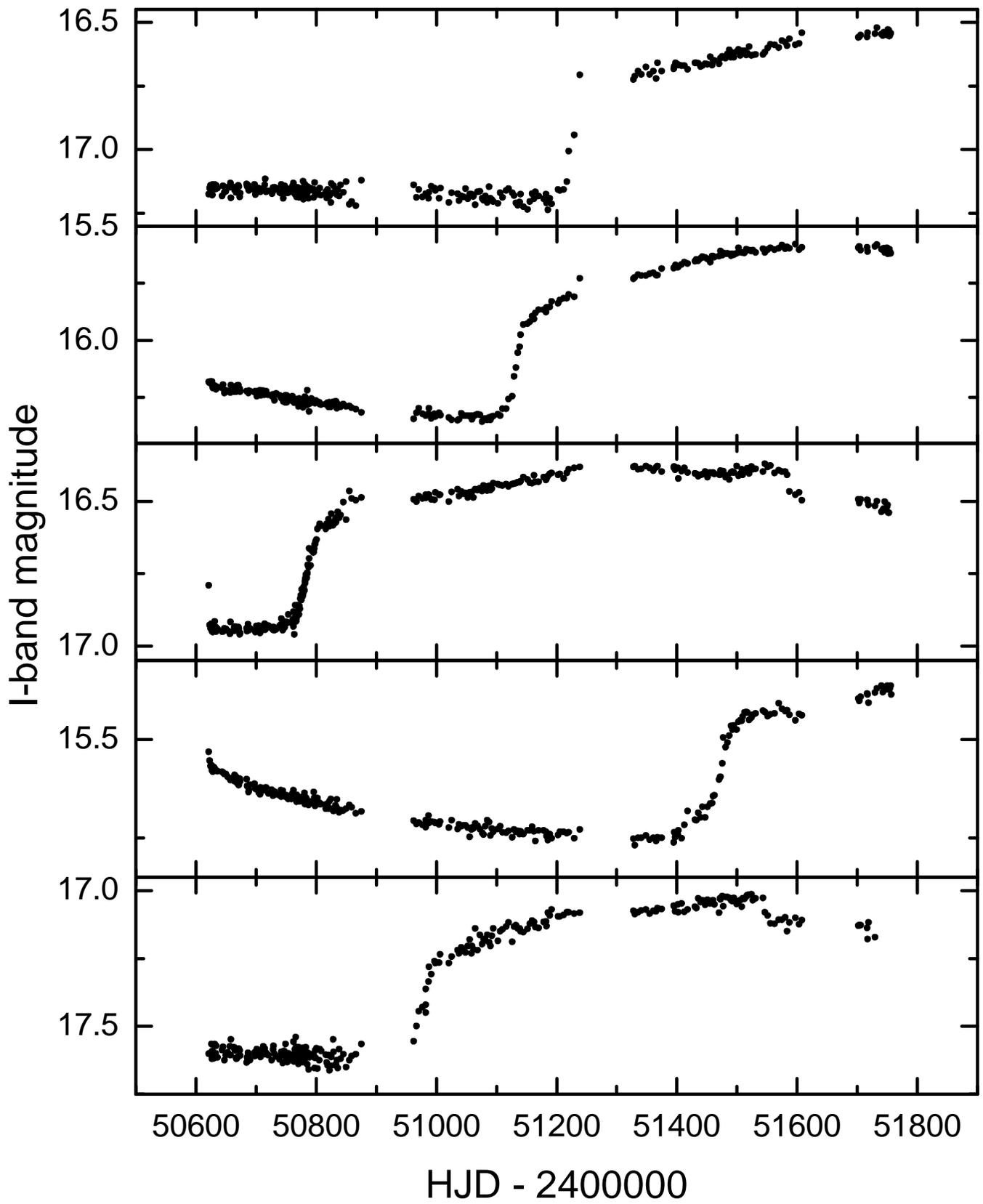
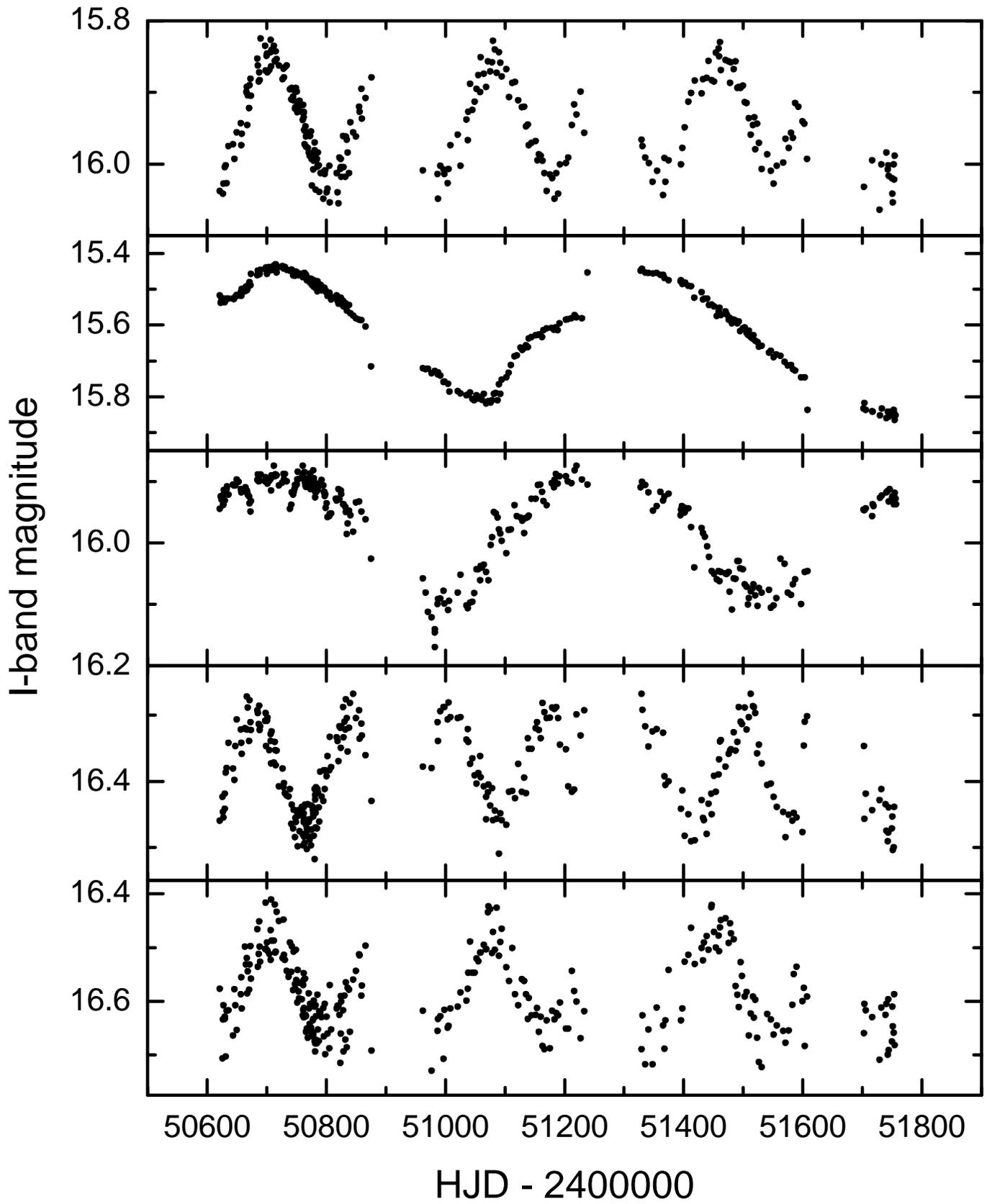
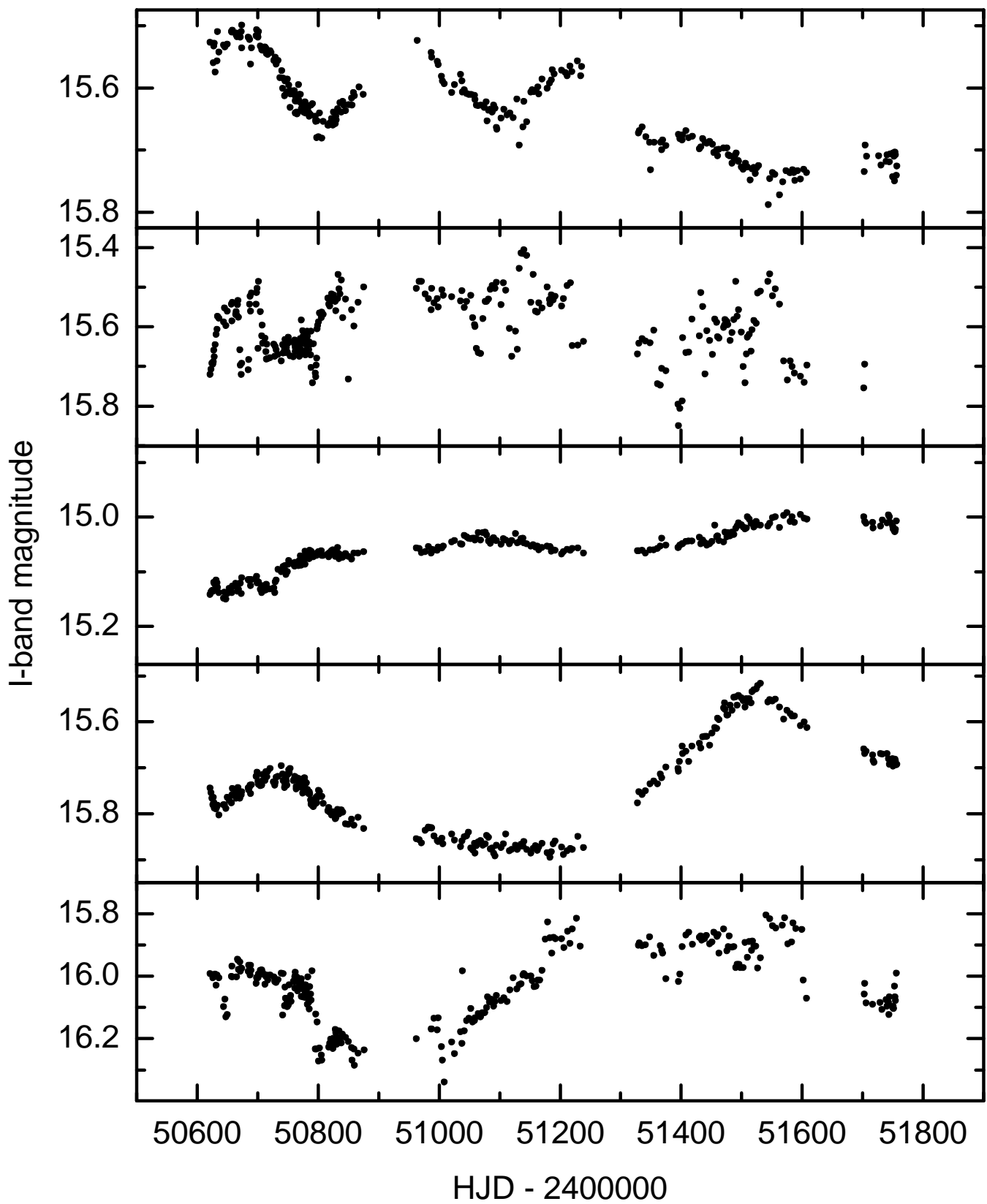
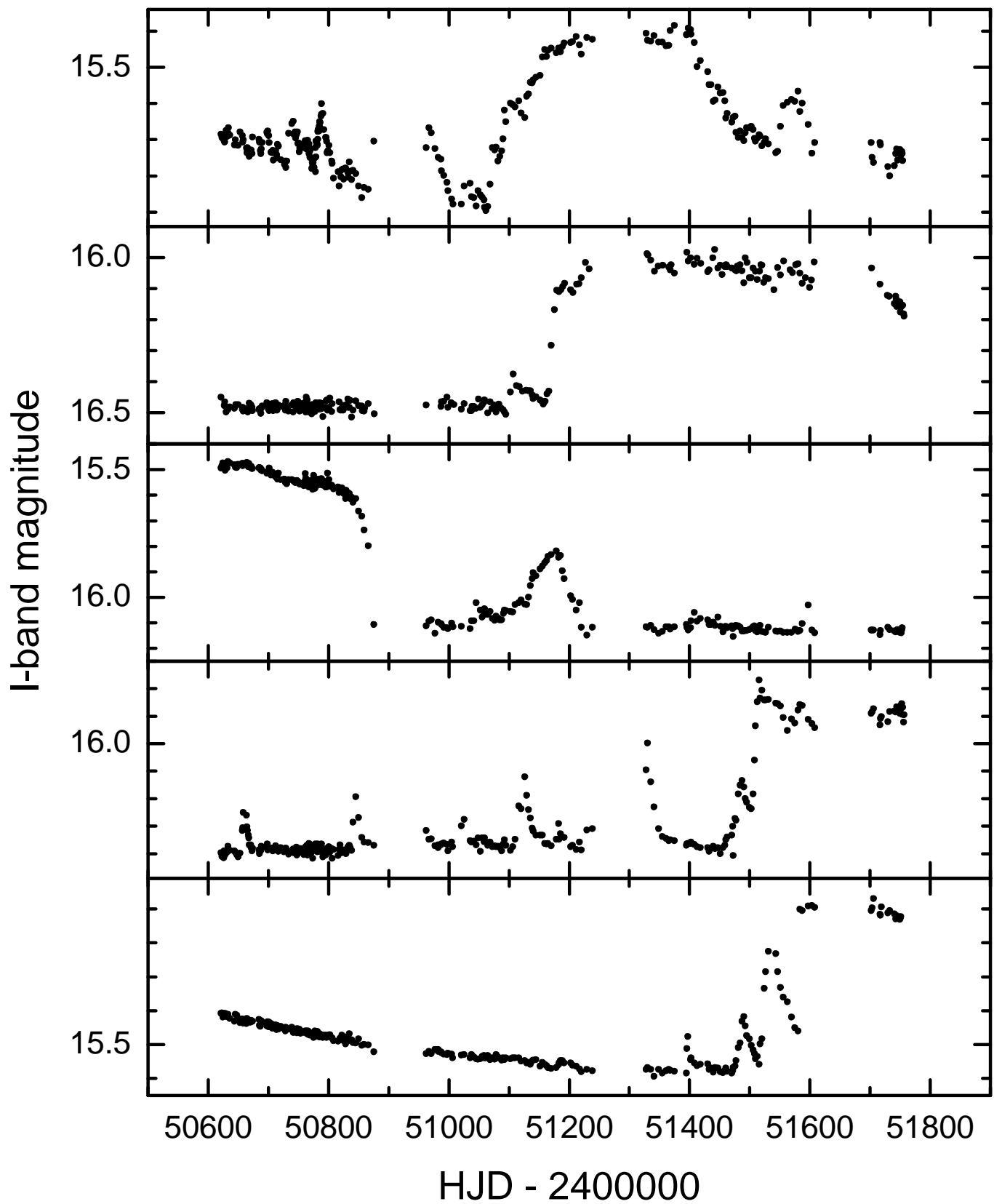
\end{document}